\documentclass[pra,superscriptaddress,nofootinbib]{revtex4}

\usepackage{amsmath}
\usepackage{blindtext}
\usepackage{algorithm}
\usepackage{algorithmic}
\usepackage{amssymb}
\usepackage{mathtools}
\usepackage{graphicx}
\usepackage{color}
\usepackage{array}
\usepackage[makeroom]{cancel}
\usepackage{bbm}
\usepackage{changes}
\usepackage{ulem}

\usepackage{tikz}
\usetikzlibrary{decorations.pathreplacing}
\usetikzlibrary{shapes,arrows,shadows,calc,patterns}

\newtheorem{defn}{Definition}
\newtheorem{lemma}{Lemma}

\tikzset{
    state/.style={
           rectangle,
           rounded corners,
           draw=black, very thick,
           minimum height=2em,
           inner sep=2pt,
           },
}

\def\red#1{\textcolor{red}{#1}}
\def\blue#1{\textcolor{blue}{#1}}

\def\ket#1{\left| #1 \right\rangle}
\def\bra#1{\left\langle #1 \right|}
\newcommand{\id}{\ensuremath{\mathbbm 1}}
\newcommand{\ketbra}[2]{\ensuremath{|#1\rangle\langle #2|}}
\newcommand{\trace}{\text{Tr}}
\newcommand{\beq}{\begin{equation}}
\newcommand{\eeq}{\end{equation}}

\begin{document}
\title{\textbf{Quantum money with nearly optimal error tolerance}}

\author{Ryan Amiri}
\affiliation{SUPA, Institute of Photonics and Quantum Sciences,
Heriot-Watt University, Edinburgh, EH14 4AS, UK}
\author{Juan Miguel Arrazola}
\affiliation{Centre for Quantum Technologies, National University of Singapore, 3 Science Drive 2, Singapore 117543}

\date{\today}
\begin{abstract}
We present a family of quantum money schemes with classical verification which display a number of benefits over previous proposals. Our schemes are based on hidden matching quantum retrieval games and they tolerate noise up to $23\%$, which we conjecture reaches $25\%$ asymptotically as the dimension of the underlying hidden matching states is increased. Furthermore, we prove that $25\%$ is the maximum tolerable noise for a wide class of quantum money schemes with classical verification, meaning our schemes are almost optimally noise tolerant. We use methods in semi-definite programming to prove security in a substantially different manner to previous proposals, leading to two main advantages: first, coin verification involves only a constant number of states (with respect to coin size), thereby allowing for smaller coins; second, the re-usability of coins within our scheme grows linearly with the size of the coin, which is known to be optimal. Lastly, we suggest methods by which the coins in our protocol could be implemented using weak coherent states and verified using existing experimental techniques, even in the presence of detector inefficiencies. 
\end{abstract}
\maketitle

\section{Introduction}

\normalem

Quantum cryptography has traditionally been associated exclusively with quantum key distribution \cite{BB84}, but it encompasses a much larger class of tasks and protocols \cite{Broadbent2016}. Notable examples are quantum signature schemes \cite{AWKA2016,AWA2015,amiri2015unconditionally}, two-party quantum cryptography \cite{lunghi2013exp,erven2014exp,ng2012experimental}, delegated quantum computation \cite{broadbent2009universal,barz2012demonstration}, covert quantum communication and steganography \cite{sanguinetti2016perfectly,bash2015quantum,arrazola2016covert,bradler2016absolutely}, quantum random number generation \cite{ma2016qrng,sanguinetti2014quantum,lunghi2015self}, quantum fingerprinting \cite{QuantumFingerprinting,arrazolaqfp,xu2015experimental,GX16}, and quantum money \cite{W1983,Gav2012,GK2015}. Historically, many of these protocols have been extremely challenging to implement with available technologies, but we are currently approaching a point where both theoretical and experimental developments have made it possible for the first experimental demonstrations to emerge. We are thus entering an exciting stage where practical quantum cryptography has begun to expand rapidly beyond the realms of quantum key distribution.

Quantum money, which was first suggested by Weisner in 1970 \cite{W1983} as a means to create money that is physically impossible to counterfeit, is one of the first examples of quantum cryptography. The basic aim of any quantum money scheme is to enable a trusted authority, the bank, to provide untrusted users with finitely re-usable, verifiable coins that cannot be forged. Verifiability ensures that honest users can prove the money they hold is genuine, while unforgeability restricts the ability of an adversary to dishonestly fabricate additional coins. Potential drawbacks of Weisner's original scheme were that verification required quantum communication between the holder and the bank, and moreover security of the scheme had not been proved rigorously. Indeed, it was shown in Refs. \cite{A09,L2010} that many variants of the scheme were vulnerable to so-called ``adaptive attacks" -- attacks in which the adversary is allowed a number of auxiliary interactions with the bank before trying to forge a coin.

In 2012, Gavinsky \cite{Gav2012} addressed both issues and presented a fully secure quantum money scheme in which coins are verified using three rounds of \emph{classical} communication between the holder of the coin and the bank. The scheme was based on hidden matching quantum retrieval games (QRGs), first introduced in Ref. \cite{YJK2004}. Nevertheless, the scheme could not be considered \emph{practical}, as the security analysis did not include the effects of noise. This issue was addressed by Pastawski et al. \cite{PYJ2011}, in which a noise tolerant quantum money scheme with classical verification was proposed that remains secure as long as the noise is less than $\frac{1}{2}-\frac{1}{\sqrt{8}}\approx 14.6\%$. The scheme requires only two rounds of communication for verification and is secure even against adaptive attacks. Following this, Ref. \cite{GK2015} presented a simpler protocol, again based on hidden matching QRGs, in which the verification procedure contained only a single round of communication, and could tolerate up to $12.5\%$ noise. 

Beyond the secret-key quantum money schemes discussed above, there has also been significant interest in public-key quantum money schemes, first proposed in \cite{A09}, offering computational security against quantum adversaries.
Since then, Farhi et al. \cite{FGH10} introduced the concepts of quantum state restoration and single-copy tomography to further rule out a large class of seemingly promising schemes. Following this result, Farhi et al. \cite{FGH12} suggested a scheme based on knot theory and conjectured that it is secure against computationally bounded adversaries. However, whether a secure public-key quantum money scheme exists without the use of oracles is an open question and, so far, the majority of schemes that were proposed have subsequently been broken \cite{LAF10}.

In this work, we focus on secret-key quantum money schemes with classical verification and propose a new scheme based on hidden matching QRGs. Utilising semi-definite programming, we provide a full security proof of our scheme, and show that by increasing the dimension of the underlying states, we can increase the error tolerance to as much as $23.03\%$ for states of dimension $n=14$, while also proving that the maximum noise tolerance in that case is $23.3\%$. Thus, the error tolerance of our protocols is nearly optimal.  We conjecture that for large dimension, the error tolerance of our protocols approaches $25\%$ asymptotically, and we further prove that $25\%$ is the maximum possible error tolerance for a wide range of quantum money protocols, including all those based on hidden matching QRGs. Increasing the error tolerance has a twofold benefit: as well as allowing the protocol to be performed in regions of higher noise than was previously possible, it also increases protocol efficiency since we show that security relies on the size of the gap between the expected error rate and the maximum tolerable error rate of the scheme, thereby allowing smaller coins. Finally, we discuss how our schemes can be implemented in practice using a coherent state encoding, while also showing that they remain secure even in the presence of limited detection efficiency. 

\subsection{Definitions and Previous Results}

In this section we state various definitions that are needed to introduce our quantum money schemes. We consider the case of quantum money ``mini-schemes" in which the bank creates only a single quantum coin and the adversary attempts to use this coin to forge another copy. It has been shown in Ref. \cite{AC2012} that by adding a classical serial number to each coin, a secure full quantum money scheme can be created directly from the secure mini-scheme, and so the two are essentially equivalent.

\begin{defn}\label{Def:QMoney}
A quantum money mini-scheme with classical verification consists of an algorithm, \emph{Bank}, which creates a quantum coin $\$$ and a verification protocol \emph{Ver}, which is a classical protocol run between a holder $H$ of $\$$ and the bank $B$, designed to verify the authenticity of the coin. The final output of this protocol is a bit $b \in\{0, 1\}$ sent by the bank, which corresponds to whether the coin is valid or not. Denote by $\text{\emph{Ver}}^B_H(\$)$ this final bit. The scheme must satisfy two properties to be secure:
\begin{itemize}
\item Correctness: The scheme is $\epsilon$-correct if for every honest holder, we have $$\text{\emph{Pr}}[\text{\emph{Ver}}^B_H(\$) = 1] \geq 1 - \epsilon.$$
\item Unforgeability: Coins in the scheme are $\epsilon$-unforgeable if for any quantum adversary who has interacted a finite and bounded number of times with the bank and holds a valid coin $\$$, the probability that she can produce two coins $\$_1$ and $\$_2$ that are verified by an honest user satisfies $$ \text{\emph{Pr}}\left[\text{\emph{Ver}}^B_H(\$_1) = 1 \wedge \text{\emph{Ver}}^B_H(\$_2) = 1 \right]\leq \epsilon,$$ where $H$ is any honest holder.
\end{itemize}
\end{defn}
The first property guarantees that all honest participants can prove the coins they own are valid, while the second property guarantees that a dishonest adversary cannot forge the coins. The definition covers adaptive attacks by allowing the adversary to interact with the bank (via the verification procedure) a finite number of times before attempting to forge the coin.

\begin{figure}[h] 
\includegraphics[width=1\textwidth]{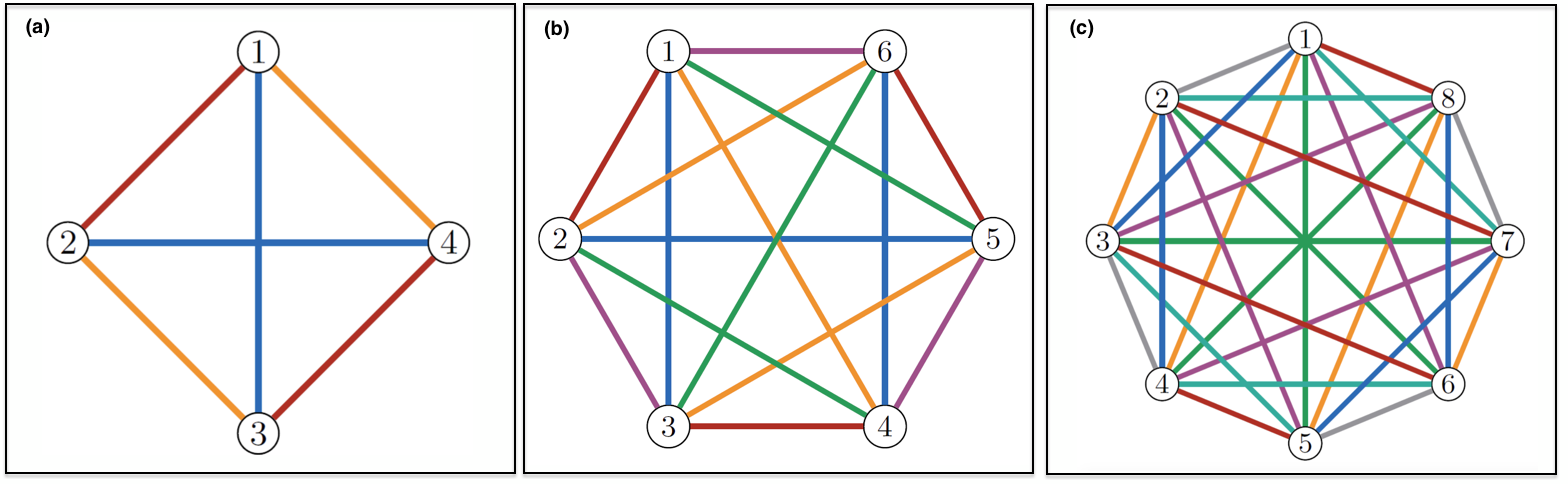}
\caption{Maximal pairwise disjoint set of matchings for (a) $n=4$, (b) $n=6$ and (c) $n=8$. Colour is used to represent each matching within the maximal pairwise disjoint set.}\label{fig:matching}
\centering
\end{figure}

The schemes presented in this paper are based on quantum retrieval games (QRGs), which we have mentioned but not formally introduced. A QRG is a protocol performed between two parties, Alice and Bob, and can be seen as a generalisation of state discrimination. Alice holds an $n$-bit string $x$, selected at random according to a probability distribution $p(x)$, which she encodes into a quantum state $\rho_x$. She sends the state to Bob, whose goal is to provide a correct answer to a given question about $x$. Mathematically, a question is modelled as a relation: if $X$ is the set of possible values $x$ can take, and if $A$ is the set of possible answers, the relation $\sigma$ is a subset of $X \times A$. If $(x,a) \in \sigma$, this means that, given $x$, the answer $a$ is a correct answer to the ``question" $\sigma$. Formally, a quantum retrieval game is defined as follows.
\begin{defn}
Let $X$ and $A$ be the sets of inputs and answers respectively. Let $\sigma \subset X\times A$ be a relation and $\{p(x), \rho_x\}$ an ensemble of states and their a priori probabilities. Then the tuple $G = (X, A, \{p(x), \rho_x\}, \sigma)$ is called a quantum retrieval game. If Bob may choose to find an answer to one of a finite number of distinct relations $\sigma_1, ..., \sigma_k$, then we write the game as $G = (X, A, \{p(x), \rho_x\}, \sigma_1, ..., \sigma_k)$.
\end{defn}
A particularly useful class of QRGs are the \emph{hidden matching} QRGs \cite{JM,GK2015,Gav2012}, in which the relations are defined by matchings. A matching $M$ on the set $[n] := \{1,2,...,n\}$, where $n$ is an even number, is a partitioning of the set into $n/2$ disjoint pairs of numbers\footnote{More precisely, this is actually the definition of a \emph{perfect} matching.}. A matching can be visualised as a graph with $n$ nodes, where edges define the elements in the matching, as illustrated in Fig. \ref{fig:matching}. In general, there are $1\times 3\times \ldots \times (n-1)=(n-1)!!$ distinct matchings of any set containing $n$ elements. For our purposes, we focus on sets of matchings where no two matchings in the set contain a common element. We call such sets \textit{pairwise disjoint}. The maximum number of pairwise disjoint matchings is $n-1$, since if we consider the element $1 \in [n]$, it must be paired in each matching with a distinct integer less than or equal to $n$. 
\begin{defn}
A maximal pairwise disjoint set of matchings, $\mathcal{R}$, is a set of pairwise disjoint matchings on $[n]$ such that $|\mathcal{R}| := n-1$.
\end{defn}
A matching on the set $[n]$ can be equivalently represented as a graph with $n$ nodes, with each each element $(i,j)$ of the matching identified with an edge in the graph. Maximal pairwise disjoint sets of matchings for $n=4,6,$ and 8 are illustrated in Fig. \ref{fig:matching}. 

In hidden matching QRGs the set of possible inputs is the set of all $n$-bit strings, each chosen with equal probability, where $n$ is an even number. Alice encodes her input into the $n$-dimensional pure state
\beq \label{eq:HM}
|\phi_x \rangle = \frac{1}{\sqrt{n}} \sum^n_{i=1} (-1)^{x_i}|i\rangle
\eeq
where $x_i$ is the $i$-th bit of the string $x$. The relations in this game are defined by the matchings: given a matching, the correct answers are the ones which correctly identify the parity of the bits connected by an edge in the matching. For example, if $(1,2)$ is an element of the matching, the measurement should output $x_1\oplus x_2$. Formally, given a perfect matching $M_1$, the set of answers is given by $$A = \big\{ (i,j,b) : i,j \in \{1,...,n\}, b\in \{0,1\} \big\}$$ and the corresponding relation is $$\sigma_1 = \{ (x, i, j, b) : x_i\oplus x_j = b \text{ and } (i,j) \in M_1 \}.$$ 
Bob is able to find a correct answer to any matching of his choice with certainty simply by measuring in the basis 
\beq \label{eq:basis}
\mathcal{B} = \{ \frac{1}{\sqrt{2}} ( |i\rangle \pm |j\rangle)\}, \:\:\:\: \text{ with } (i,j)\in M.
\eeq
This is because the outcome $\frac{1}{\sqrt{2}} ( |i\rangle + |j\rangle)$ can only occur if $x_i\oplus x_j = 0$, and similarly $\frac{1}{\sqrt{2}} ( |i\rangle - |j\rangle)$ can only occur if $x_i\oplus x_j = 1$.

Previous quantum money schemes based on hidden matching QRGs have used only two matchings for verification. In the following section, we generalise these schemes to the case of an arbitrary number of matchings and show that this allows us to significantly increase the noise tolerance of the resulting schemes. 

%%%%%%%%%%%%%%%%%%%%%%%%%%%%%%%%%%%%%%%%%%%%%%%%%%%%%%%%%%%%%%%%%%%%%%%%%%%
%%%%%%%%%%%%%%%%%%%%%%%%%%%%%%%%%%%%%%%%%%%%%%%%%%%%%%%%%%%%%%%%%%%%%%%%%%%
%%%%%%%%%%%%%%%%%%%%%%%%%%%%%%%%%%%%%%%%%%%%%%%%%%%%%%%%%%%%%%%%%%%%%%%%%%%

\section{Quantum money scheme} \label{sec:scheme}

Here we present a quantum money scheme which is secure even in the presence of up to $23\%$ noise. As in Ref. \cite{GK2015}, the verification protocol requires only one round of classical communication. 

\begin{figure}[h!] 
\centering
\begin{tikzpicture}
\node[state] (Bank) at (0,0)
 {\begin{tabular}{ll}
  \textbf{Choose:} & \textbf{Initialise:}\\
  $x_1 = 01011011$, & $\:\:r=0^q,$\\
  $x_2 = 11000010$, & $\:\:s=0.$\\
  $\:\vdots$ & \\
  $x_q = 10101110$. & 
  \end{tabular}
  };
\node [above] (Bankpic) at (Bank.north)
    {\includegraphics[height=1.4cm, width=.3\textwidth]{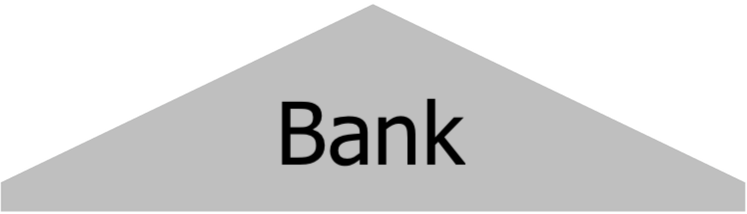}};
\node [below] (base) at (Bank.south)
    {\includegraphics[width=.3\textwidth]{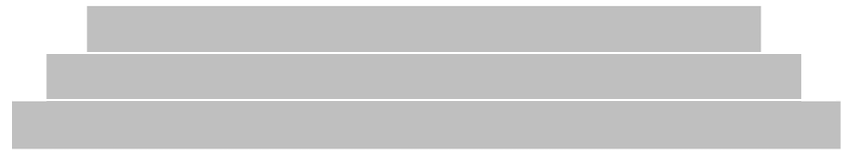}};
\node (Holder) at (12.5,0.3)
    {\includegraphics[width=.15\textwidth]{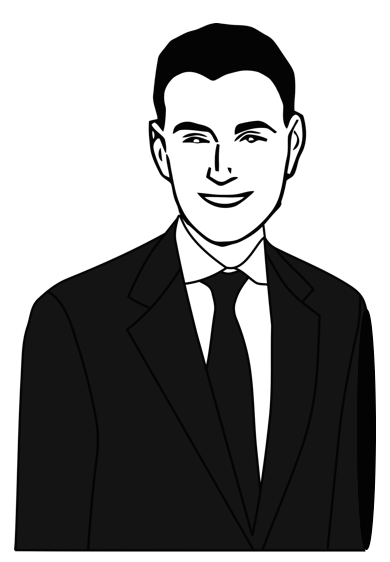}};
\draw[->,thick] (Bank.east) -- ($(Holder.west)!(Bank)!(Holder.west)$)
    node[midway,fill=white, above] {$(\$, r) = (\rho_{x_1} \otimes \dots \otimes \rho_{x_q}, r)$};
\end{tikzpicture}
\caption{Schematic illustration of the Bank algorithm for $n=8$. The bank selects $q$ $8$-bit strings and initialises the $q$-bit register $r$ to the zero string. The bank creates the corresponding hidden matching states and sends these, together with $r$, to the holder of the coin.}\label{fig:bank}
\end{figure}

\begin{figure}[h!] 
\centering
\begin{tikzpicture}
\node [state] (verify) at (0,0)
    {\renewcommand{\arraystretch}{1.2}
    \begin{tabular}{cccccccccc}
  $r\: :$ & $0$ & $0$ & $\red{1}$ & $0$ & $\red{1}$ & $ 0$ & $ 0$ & $ 0$ & $\dots $ \\
  $\$\: :$ & $\blue{\rho_{x_1}} $ & $ \rho_{x_2} $ & $ \red{\xcancel{\rho_{x_3}}}$ & $\blue{\rho_{x_4}}$ & $\red{\xcancel{\rho_{x_5}}}$ &  $ \rho_{x_6} $ & $ \blue{\rho_{x_7}}$ & $ \blue{\rho_{x_8}}$ & $ \dots $ \\
  M : & $\blue{\sigma^\prime_1} $ & - & - & $\blue{\sigma^\prime_4}$ & -  & - & $\blue{\sigma^\prime_7}$ &$\blue{\sigma^\prime_8}$ & $ \dots $ \\
  $\rightarrow$ & $\blue{d_1} $ & - & - & $\blue{d_4}$ & - & - & $\blue{d_7}$ & $\blue{d_8}$ & $ \dots $\\
  \end{tabular}
  };
\node [above] (Holder) at (verify.north)
    {\includegraphics[width=.12\textwidth]{Holder2.png}};
\node [state] (check) at (10,0)
    {\renewcommand{\arraystretch}{1.2}
    \begin{tabular}{cccccc}
  $\longrightarrow$ & $(1, \sigma^\prime_1, d_1)$ & $(4, \sigma^\prime_4, d_4)$ & $(7, \sigma^\prime_7, d_7)$ &$(8, \sigma^\prime_8, d_8)$ & $ \dots $ \\
  $x \: :$ & $x_1 $ & $x_4$ & $x_7$ & $x_8$ & $ \dots $ \\
  Outcome: & $ \checkmark/\times$  & $ \checkmark/\times$ & $ \checkmark/\times$ & $\checkmark/\times$ & $ \dots $\\
   & $s \rightarrow s+1$ & & & &
  \end{tabular}
  };
\node [above] (Bank) at (check.north)
    {\includegraphics[width=.13\textwidth]{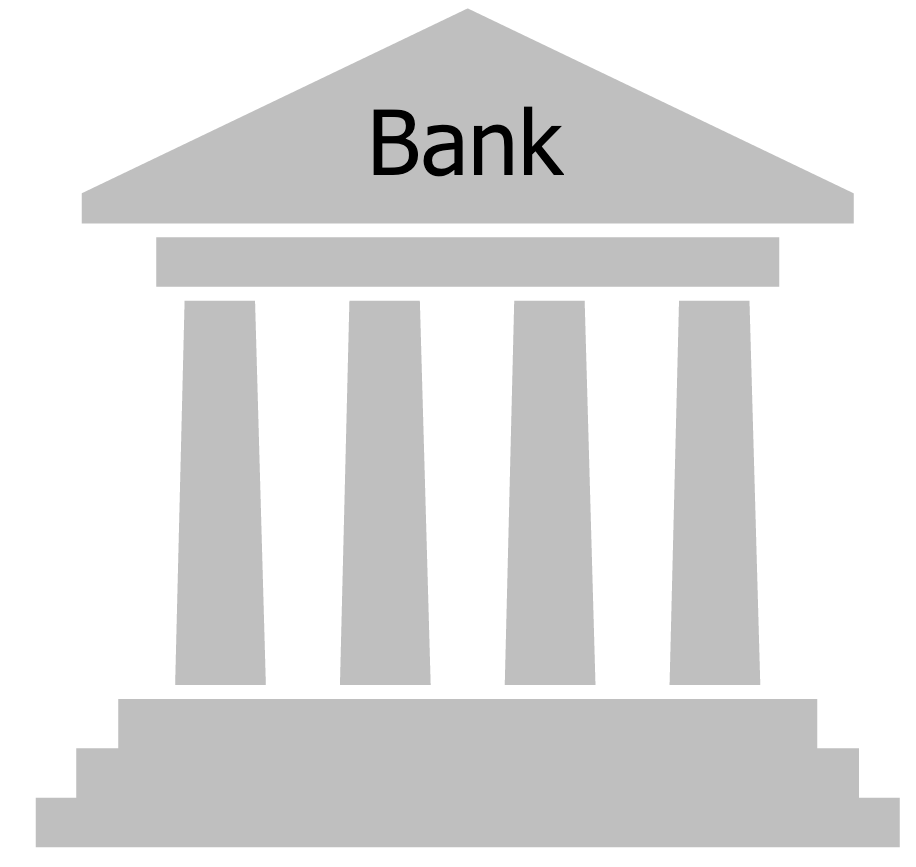}};
\draw[->,thick] (verify.east) -- ($(check.west)!(verify)!(check.west)$)
    node[midway,fill=white, above] {$\{(i, \sigma^\prime_i, d_i)\}$};
\end{tikzpicture}
\caption{Schematic showing the verification algorithm. The verifier selects a sample $\{\rho_{x_1},\rho_{x_4},\rho_{x_7},\rho_{x_8},\ldots\}$ of the states contained within the coin which have an $r$ value of $0$. He randomly chooses matching measurements and applies them to get classical measurement outcomes which he sends to the bank, together with the index of the state and the matching chosen. The bank checks these against its secret strings, as well as checking $s<T$. Finally, the bank declares an output based on the number of incorrect outcomes.}\label{fig:ver}
\end{figure}

In this scheme, the bank randomly chooses a number of $n$-bit classical strings and encodes each of them into the hidden matching states, given by Eq. \eqref{eq:HM}. Essentially, the coin is a collection of these independent quantum states, and each of the quantum states can be thought of as an instance of a QRG. We assume that there is a maximal pairwise disjoint set of matchings on $[n]$, known to all participants, which we call $\mathcal{R}$. This set specifies the $n-1$ possible relations defined within each QRG, and each state in the coin represents a QRG. To verify a coin, the holder will pick a small selection of the states from the coin and randomly choose a relation for each. The holder will perform the appropriate measurement (defined by Eq. \eqref{eq:basis}) to get an answer for each QRG under each chosen relation. The holder then sends these answers to the bank which returns whether more than a specified fraction of the answers are correct or not. If they are, the coin is accepted as valid; otherwise, it is rejected. The scheme is formally defined below and illustrated in Figs. \ref{fig:bank} and \ref{fig:ver}.

\begin{algorithm}[H]
\floatname{algorithm}{Bank Algorithm}
\renewcommand{\thealgorithm}{}
\caption{}
\begin{algorithmic}[1]
\STATE The bank independently and randomly chooses $q$ $n$-bit strings which we will call $x^{1}, ..., x^{q}$. 
\STATE For $i\in [q]$, the bank creates $\phi_{x^i} := |\phi_{x^i}\rangle\langle \phi_{x^i}|$, where $$|\phi_{x^i}\rangle := \frac{1}{\sqrt{n}} \sum^n_{j=1} (-1)^{x^i_j}|j\rangle.$$ For each $i$ we define the QRG $G_i = (S_i, A_i, \{\phi_{x^i} \}_{x^i}, \sigma_1, ..., \sigma_{n-1})$, where $\mathcal{R} = \{\sigma_1, ..., \sigma_{n-1}\}$ is a maximal pairwise disjoint set of matchings known to all participants in the scheme.
\STATE The bank creates the classical binary register, $r$, and initialises it to $0^q$.
\STATE The bank creates the counter variable $s$ and initialises it to $0$.
\STATE The pair $ (\$,r) = (\bigotimes_{i=1}^q \phi_{x^i}, r) $ is the coin for the mini-scheme. The bank keeps the counter $s$ in order to keep track of the number of verification attempts.
\end{algorithmic}
\end{algorithm} 

\begin{algorithm}[H]
\floatname{algorithm}{Ver Algorithm}
\renewcommand{\thealgorithm}{}
\caption{}
\begin{algorithmic}[1]
\STATE The holder of the coin randomly chooses a subset of indices, $L\subset [q]$ such that $r_i =0$ for each $i \in L$. The indices $i\in L$ specify the selection of games $G_i$ which will be used as tests in the verification procedure. For each $i\in L$, the holder sets the corresponding bit of $r$ to be $1$ so that this game cannot be used in future verifications.
\STATE For each $i\in L$, the holder picks a relation $\sigma^\prime_i$ at random from $\mathcal{R}$ and applies the appropriate measurement to obtain outcome $d_i$.
\STATE The holder sends all triplets $(i, \sigma^\prime_i, d_i)$ to the bank.
\STATE The bank checks that $s<T$, where $T$ is the pre-defined maximum number of allowed verifications for the coin. If $s = T$, the bank declares the coin as invalid.
\STATE For each $i$, the bank checks whether the answer is correct by comparing $(i, \sigma^\prime_i, d_i)$ to the secret $x^i$ values. The bank accepts the coin as valid if and only if more than $l(c-\delta)$ of the answers are correct, where $c$ is a correctness parameter of the protocol, $l = |L|$, and $\delta$ is a small positive constant.
\STATE The bank updates $s$ to $s+1$.
\end{algorithmic}
\end{algorithm} 

We say that an instance of the verification algorithm has been passed/failed if the final output by the bank is ``valid"/``invalid" respectively. Coins can be verified at most $T$ times until the Hamming weight of $r$ is greater than $Tl$, at which point the coin is returned to the bank to be refreshed. We choose $T$ to be small but linear in $q$. Any such choice would be acceptable but, for the sake of definiteness, in what follows we set $T := q/(1000l)$. We note that having $T$ scale linearly with $q$ is optimal for any quantum money scheme \cite{Gav2012} and that this is an improvement over previous protocols (for example those in Refs. \cite{GK2015, Gav2012}). 

The parameter $c$ represents the probability that an honest verifier obtains a correct outcome for a QRG in an honest run of the protocol. In the ideal setting $c=1$, since an honest participant in possession of a correct state will always be able to get a correct answer to a relation. Of course, in practice system imperfections inevitably lead to errors so that even when all participants are honest, it is not certain that the holder's measurement will return a correct answer. Thus, in the presence of errors, we must have $c<1$, and the smallest value of $c$ for which we can retain security determines the noise tolerance of the protocol.

We note that this scheme requires the bank to maintain a small classical database to record the number of times the verification protocol has been run -- i.e. the bank's database is ``non-static", and must be updated after each run of verification. Although this requirement demands more from the bank than completely static database models, we believe the requirement is both minimal and realistic, and allows significant simplifications to the security analysis. Nevertheless, in some cases it may be desirable for the bank to have a completely static database -- for example in applications in which the bank consists of many small, decentralised branches wary of attacks spanning multiple bank locations. In this case, by adding an additional round of classical communication in the verification protocol, our scheme can be transformed into a fully static database scheme which retains the same level of noise tolerance. Security can be proved by directly applying the arguments in Ref. \cite{Gav2012} to show that the additional verification attempts do not (significantly) help the adversary\footnote{We are able to apply the arguments in Ref. \cite{Gav2012} because, although our scheme uses more than two matchings, when taken pairwise any two matchings within our scheme are independent.}.
%%%%%%%%%%%%%%%%%%%%%%%%%%%%%%%%%%%%%%%%%%%%%%%%%%%%%%%%%%%%%%%%%%%%%%%%%%%%%%%%%%%%

\subsection{Security} \label{sec:security}
In this section we prove that the scheme defined above is secure according to Definition \ref{Def:QMoney}. 
\subsubsection{Correctness}
Correctness of the scheme follows simply from the Hoeffding bound \cite{Hoeffding}. In the honest case, if the holder of a coin has probability $c$ of getting a correct answer for each of the $l$ QRGs selected in the verification protocol, then his probability of getting fewer than $(c-\delta)l$ correct answers overall is bounded by
\beq
\mathbb{P}(\text{Honest Fail}) \leq e^{-2l\delta^2}.
\eeq
Based on the security analysis in the following section, we choose $\delta$ to be half of the gap between the error rate an honest participant expects and the minimum error rate the adversary can achieve. I.e. we set $\delta := (e_{\text{min}}-\beta)/2$, where $e_{\text{min}}$ is the minimum error rate achievable by the adversary (derived below in Eq. \eqref{eq:perror}), and $\beta := 1-c$ is the error rate expected in an honest run of the protocol.

\subsubsection{Unforgeability}
We assume the adversary is in possession of a valid coin and first address a simple forging strategy available to the adversary based on manipulating the $r$ register attached to the coin. The adversary is allowed to set at most $q/1000$ of the $r$ register entries to $1$. She creates $(\$_1,r_1)$ and $(\$_2,r_2)$ to send to the two honest verifiers, Ver$_1$ and Ver$_2$ respectively. If she sets $r_1(i)=1$ and $r_2(i)=0$, she can be certain that Ver$_1$ will not select the $i$'th state to test, and so can forward the perfect state to Ver$_2$. In this way, $q/1000$ of the states in the coins sent to each verifier will be perfect, and will not cause errors. The remaining positions must have $r$ register values of $0$ for both verifiers. Similarly, the adversary is able to use the auxiliary verification attempts to her advantage. We make a worst-case assumption and assume that the adversary gets full knowledge of every state used in an auxiliary verification attempt. Since there are at most $T$ attempts allowed, each of which involve $l$ states, the adversary knows the identity of at most $q/1000$ of the states. Since the states are prepared independently, this knowledge does not provide any information on the remaining states. 

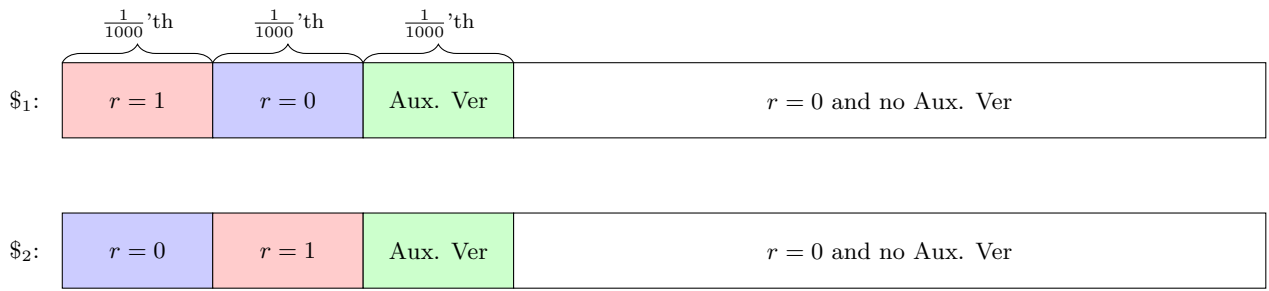
\begin{figure}[H]
\centering
\begin{tikzpicture}[scale=1]
\draw[step=1cm,white,very thin] (-1,0) grid (16,-4);
\filldraw[fill=white, draw=white] (-1,-1) rectangle node{$\$_1$:} (0,-2);
\filldraw[fill=red!20!white, draw=black] (0,-1) rectangle node{$r=1$} (2,-2);
\filldraw[fill=blue!20!white, draw=black] (2,-1) rectangle node{$r=0$} (4,-2);
\filldraw[fill=green!20!white, draw=black] (4,-1) rectangle node{Aux. Ver} (6,-2);
\filldraw[fill=white, draw=black] (6,-1) rectangle node{$r=0$ and no Aux. Ver} (16,-2);

\draw [decorate,decoration={brace,amplitude=7pt},xshift=0pt,yshift=0pt]
(0,-1) -- (2,-1) node [black,midway, yshift=15pt] 
{\footnotesize $\frac{1}{1000}$'th};
\draw [decorate,decoration={brace,amplitude=7pt},xshift=0pt,yshift=0pt]
(2,-1) -- (4,-1) node [black,midway, yshift=15pt] 
{\footnotesize $\frac{1}{1000}$'th};
\draw [decorate,decoration={brace,amplitude=7pt},xshift=0pt,yshift=0pt]
(4,-1) -- (6,-1) node [black,midway, yshift=15pt] 
{\footnotesize $\frac{1}{1000}$'th};

\filldraw[fill=white, draw=white] (-1,-3) rectangle node{$\$_2$:} (0,-4);
\filldraw[fill=blue!20!white, draw=black] (0,-3) rectangle node{$r=0$} (2,-4);
\filldraw[fill=red!20!white, draw=black] (2,-3) rectangle node{$r=1$} (4,-4);
\filldraw[fill=green!20!white, draw=black] (4,-3) rectangle node{Aux. Ver} (6,-4);
\filldraw[fill=white, draw=black] (6,-3) rectangle node{$r=0$ and no Aux. Ver} (16,-4);

\end{tikzpicture}

\caption{Representation of the states within the quantum coins sent to the verifiers. The first block on the far left represents all states for which the adversary set $r=1$ for Ver$_1$, and $r=0$ for Ver$_2$. The adversary knows that Ver$_1$ cannot select these states for testing, and so is able to forward on the perfect states to Ver$_2$. The second block of states represents the same, but with the roles of the verifiers reversed. The Aux. Ver states in the diagram are the ones that we assume are known to the adversary via auxiliary verifications. The remaining states in white are the ones we consider below -- those states for which the $r$ register is zero for both verifiers, and which have not been used in auxiliary verifications. } \label{fig:forge}
\end{figure}

The combined effect of the above two strategies is that the adversary is able to exactly replicate $q/500$ of the states in the coin, as shown in Fig. \ref{fig:forge}. To prove coins are unforgeable, we consider the remaining $997q/1000$ states for which the $r$ register is zero for both verifiers, and for which the adversary has no auxiliary information. In reference to Fig. \ref{fig:forge}, we refer to these states as the white states, and start by considering a single such state, $\phi_{x^i}:= \ket{\phi_{x^i}}\bra{\phi_{x^i}}$, contained in the coin. For simplicity, we drop the superscript on the $n$-bit strings $x^i$ in all that follows.

The idea behind the proof is to relate the probability that the forger can use a single white state to create two states that pass the verification test of the two honest verifiers, to the average fidelity of these two states with the original state $\ket{\phi_x}$. The maximisation of this average fidelity corresponds to the optimal attack, which can be cast as a semi-definite program. By focusing on the dual program, we can upper bound the value of the semi-definite program and therefore bound the forging probability of the adversary. Lastly, we show that coherent attacks on multiple states cannot help the adversary to forge.

Since the adversary has a valid coin, she holds the unknown state
\beq \label{eq:start}
|\phi_x\rangle = \frac{1}{\sqrt{n}} \sum^n_{i=1} (-1)^{x_i}|i\rangle.
\eeq
From this state, the adversary wishes to create two states, $\eta_x$ and $\tau_x$, which, when measured by the honest verifiers, will give the correct answer to a randomly chosen relation in $\mathcal{R}$. Consider the normalised state sent to $\textrm{Ver}_1$,
\beq
\eta_x = \sum^n_{i,j = 1} a_{ij}|i\rangle\langle j|.
\eeq
Suppose the verifier chooses to measure using the matching $M_\alpha = \{(i_1, j_1),...,(i_{n/2},j_{n/2})\}$, where $\alpha \in\{1,2,\ldots,n-1\}$. To find a correct answer to the relation $\sigma_\alpha$ defined by this matching, an honest verifier will apply the measurement with projectors in the set $\{\ket{+_{i_kj_k}}\bra{+_{i_kj_k}}, \ket{-_{i_kj_k}}\bra{-_{i_kj_k}} \: : \: k=1, ..., n/2\}$, where $\ket{\pm_{i_kj_k}} := \frac{1}{\sqrt{2}}(\ket{i_k} \pm \ket{j_k})$. An incorrect result is obtained whenever the verifier finds an incorrect value for $x_{i_k}\oplus x_{j_k}$, which happens whenever the measurement outcome is one of the form
\beq
\frac{1}{\sqrt{2}} ( |i\rangle - (-1)^{x_i\oplus x_j} |j\rangle ).
\eeq
This happens with probability
\beq
p^{\alpha, x}_{\text{Ver}_1} = \frac{1}{2} \left( 1 - \sum^{n/2}_{k=1}(-1)^{x_{i_k}\oplus x_{j_k}}a_{i_kj_k} + (-1)^{x_{i_k}\oplus x_{j_k}}a_{j_ki_k} \right).
\eeq
Thus, the probability of an incorrect answer to $\sigma_\alpha$ is given by a subset of the off-diagonal elements of the density matrix $\eta_x$. The off-diagonal elements occurring are exactly those with indices paired by the matching $M_\alpha$. Since the set of relations form a maximal pairwise disjoint set, the off-diagonal matrix elements appearing in the error probability for different relations will all be distinct. Therefore, averaging over all possible relations that could be chosen by the verifier allows us to significantly simplify the adversary's error probability, which becomes
\beq \label{eq:aver}
p^x_{\text{Ver}_1} = \frac{1}{n-1}\sum_{\alpha=1}^{n-1}p^{\alpha, x}_{\text{Ver}_1} = \frac{1}{2(n-1)} \left( n - \sum^n_{i, j=1} (-1)^{x_{i}\oplus x_{j}}a_{ij} \right) = \frac{n}{2(n-1)}(1-F_x),
\eeq
where we have defined
\beq
F_x := \langle \phi_x | \eta_x | \phi_x \rangle = \frac{1}{n} \sum_{i,j} (-1)^{x_{i}\oplus x_{j}}a_{ij}.
\eeq
Since the adversary does not know the secret string $x$, rather than holding the state in Eq. \eqref{eq:start}, she instead holds a mixture over the possible $x$ values. We define $F := \frac{1}{2^{n}}\sum_xF_x$ and take an average over $x$ values to get
\begin{equation} \label{eq:aver2}
p_{\text{Ver}_1} = \frac{1}{2^n} \sum_x p^x_{\text{Ver}_1} = \frac{1}{2^n}  \sum_x \frac{n}{2(n-1)} \left( 1 - F_x \right) = \frac{n}{2(n-1)} \left( 1 - F \right).
\end{equation}
Essentially then, to successfully forge a coin, the adversary is trying to create two states, $\eta_x$ and $\tau_x$, which both have a high fidelity with the original state $|\phi_x\rangle$. Let's define $G_x = \langle \phi_x| \tau_x | \phi_x \rangle$, and $G := \frac{1}{2^{n}}\sum_xG_x$. For the purpose of forging, the adversary needs \emph{both} $\textrm{Ver}_1$ and $\textrm{Ver}_2$ to accept the coin she sends, which requires her to make both error probabilities as small as possible. From the above result, we can relate this to maximising the average fidelity of the states $\eta_x$ and $\tau_x$ with the original state. This problem can be cast as a semi-definite program as follows.

Let $\Psi : L(\mathcal{X}) \rightarrow L(\mathcal{Y} \otimes \mathcal{Z} )$ be a physical channel taking states in Hilbert space $\mathcal{X}$ to states in the Hilbert space $\mathcal{Y}\otimes \mathcal{Z}$, where both $\mathcal{Y}$ and $\mathcal{Z}$ are isomorphic to $\mathcal{X}$. We want to find the channel that maximises
\beq \label{eq:sdp}
\overline{F}=\frac{1}{2^n} \sum^{2^n}_{x=1} \frac{\langle \phi_x |\eta_x|\phi_x\rangle + \langle \phi_x |\tau_x|\phi_x\rangle}{2},
\eeq
where $\eta_x = \trace_{\mathcal{Z}}\left[ \Psi(|\phi_x\rangle\langle \phi_x |)\right]$ and $\tau_x = \trace_{\mathcal{Y}}\left[ \Psi(|\phi_x\rangle\langle \phi_x |)\right]$. In other words, $\eta_x$ is the reduced state of the channel output representing the state held by $\textrm{Ver}_1$, and $\tau_x$ is the reduced state of the channel output representing the state held by $\textrm{Ver}_2$. This maximisation is subject to $\Psi$ being a completely positive trace preserving linear map. To express this maximisation in the standard form of a semi-definite program, we express the channel as an operator using the Choi representation. We fix the preferred basis to be $\{ |i\rangle \}_{i=1,...,n}$, the basis used to define the hidden matching states in the ensemble. Given this choice, the Choi operator corresponding to the channel $\Psi$ is an operator $J(\Psi)$ in $L(\mathcal{X}\otimes \mathcal{Y} \otimes \mathcal{Z})$, given by
\beq
J(\Psi) = \sum^n_{i,j = 1} |i\rangle\langle j|_{\mathcal{X}} \otimes \Psi(|i\rangle\langle j|)_{\mathcal{Y}\mathcal{Z}}
\eeq
Using the facts that $\langle \phi_x | i\rangle = \langle i | \phi_x\rangle $ for all states in the ensemble, and that $\Psi$ is a linear map, it can be shown that
\beq
\trace_{\mathcal{X}\mathcal{Y}\mathcal{Z}}\Bigg[ \Big( \phi^{\mathcal{X}}_x \otimes \phi^{\mathcal{Y}}_x \otimes \id^{\mathcal{Z}}\Big) J(\Psi)\Bigg] = \langle\phi_x |\eta_x|\phi_x\rangle_{\mathcal{Y}},
\eeq
and similarly that 
\beq
\trace_{\mathcal{X}\mathcal{Y}\mathcal{Z}}\Bigg[ \Big( \phi^{\mathcal{X}}_x \otimes \id^{\mathcal{Y}} \otimes \phi^{\mathcal{Z}}_x \Big) J(\Psi)\Bigg] = \langle\phi^x |\tau_x|\phi^x\rangle_{\mathcal{Z}},
\eeq
where here, for ease of notation, we have used the superscript to denote the relevant Hilbert space. With this we can rewrite the problem in Eq. \eqref{eq:sdp} as the problem of finding the operator $J(\Psi)$ which maximises
\beq
\frac{1}{2^{n+1}}\sum^{2^n}_{x=1} \trace_{\mathcal{X}\mathcal{Y}\mathcal{Z}}\Bigg[ \Big( (\phi^{\mathcal{X}}_x \otimes \phi^{\mathcal{Y}}_x \otimes \id^{\mathcal{Z}}) + (\phi^{\mathcal{X}}_x \otimes \id^{\mathcal{Y}} \otimes \phi^{\mathcal{Z}}_x ) \Big) J(\Psi)  \Bigg].
\eeq
The conditions that the channel must be completely positive and trace preserving lead to the conditions that $J(\Psi)$ must be positive semidefinite and $\trace_{\mathcal{Y}\mathcal{Z}}(J(\Psi)) = \id_{\mathcal{X}}$. Written in standard form, the semidefinite program corresponding to the maximum average fidelity is given by
\begin{equation}
\begin{aligned}
& \text{Maximise:}
& & \langle Q(n), X \rangle \\
& \text{subject to:}
& & \trace_{\mathcal{Y}\mathcal{Z}}(X) = \id_{\mathcal{X}} \\
&&& X \geq 0,
\end{aligned}
\end{equation}
where 
\beq
Q(n) = \frac{1}{2^{n+1}} \sum^{2^n}_{x=1}  \Big( (\phi^{\mathcal{X}}_x \otimes \phi^{\mathcal{Y}}_x \otimes \id^{\mathcal{Z}}) + (\phi^{\mathcal{X}}_x \otimes \id^{\mathcal{Y}} \otimes \phi^{\mathcal{Z}}_x ) \Big).
\eeq
The dual problem is simply
\begin{equation}
\begin{aligned}
& \text{Minimise:}
& & \trace(Y) \\
& \text{subject to:}
& & \id_{\mathcal{Y}\mathcal{Z}}\otimes Y \geq Q(n) \\
&&& Y \in \text{Herm}(\mathcal{X}),
\end{aligned}
\end{equation}
since $\langle \id_{\mathcal{X}}, Y \rangle = \trace(Y)$ and the adjoint of the partial trace is the extension by the identity. The dual problem approaches the optimal value from above, so any feasible point (i.e. any operator $Y$ that satisfies the constraints of the dual problem) gives us an upper bound on the maximum average fidelity. A feasible point can easily be found in terms of the matrix $Q(n)$ as
\beq
Y=||Q(n)||_{\infty}\id_{\mathcal{X}}
\eeq
so that we arrive at the following upper bound on the average fidelity:
\beq
\overline{F}\leq n||Q(n)||_{\infty}.
\eeq
Thus, for quantum money protocols using states of dimension $n$ and a maximal disjoint set of matchings, we can upper bound the error probability of the adversary in terms of the operator norm of $Q(n)$. Computing this norm for different values of $n$ leads to the bound
\beq
\overline{F}\leq \frac{1}{2}+\frac{1}{n}
\eeq
which we have verified numerically for $n\leq 14$ and we conjecture holds for any $n$. From now on, we simply assume that $n\leq 14$. The analysis above enables us to restrict the achievable error probabilities for the two verifiers on a single game as
\beq
\begin{rcases}
p_{\text{Ver}_1} = \frac{n}{2(n-1)} \left( 1 - F \right) \\
p_{\text{Ver}_2} = \frac{n}{2(n-1)} \left( 1 - G \right)
\end{rcases}
\quad
\text{ subject to: } \frac{1}{2} (F+G) \leq \frac{1}{2}+\frac{1}{n},
\eeq
which leads to
\beq \label{eq:bound}
p_{\text{Ver}_1} + p_{\text{Ver}_2} \geq \frac{1}{2} - \frac{1}{2(n-1)}.
\eeq
Until now, we have considered only a single white state out of the $l$ games used in the verification protocol. Let us now consider $l$ such games, and let $p^{(i)}_{\text{Ver}_j}$ be the error probability for honest verifier $j$ on the $i$'th run of the verification protocol. We claim that when we have $l$ independent white states (in the sense that each $x^i$ is chosen independently), it is still the case that
\beq
p^{(i)}_{\text{Ver}_1} + p^{(i)}_{\text{Ver}_2} \geq \frac{1}{2} - \frac{1}{2(n-1)}
\eeq
for all $i$, regardless of the outcomes of previous measurements made by the verifiers. Though intuitively reasonable, this claim is far from trivial, but can be proved using a teleportation argument due to Croke and Kent \cite{CrokeKent} (See Appendix A) so that, essentially, we can imagine the adversary acts independently on each game in the verification protocol. Therefore, on each and every white state, at least one verifier must have an error probability of at least
\beq
\frac{1}{2}(p^{(i)}_{\text{Ver}_1} + p^{(i)}_{\text{Ver}_2}) = \frac{1}{4} - \frac{1}{4(n-1)}.
\eeq
Overall, if we include the effects of $r$ register manipulation and auxiliary verifications, at least one verifier, say Ver$_1$, must have an average error probability over all $l$ games of at least
\beq \label{eq:perror}
e_{\text{min}} = \frac{997}{999}\left(\frac{1}{4} - \frac{1}{4(n-1)}\right) \approx \frac{1}{4} - \frac{1}{4(n-1)} 
\eeq
Using Hoeffding's inequality, the probability of both verifiers accepting the coin can be bounded as
\beq
\mathbb{P}(\text{Both $\textrm{Ver}_1$ and $\textrm{Ver}_2$ generate outcome ``Valid"}) \leq \mathbb{P}(\text{$\textrm{Ver}_1$ generates outcome ``Valid"}) \leq e^{-2l\delta^2},
\eeq
where $\delta = (e_{\text{min}} - \beta)/2$, as above. As long as $\beta < e_{\text{min}}$, the Hoeffding bound can be used to show that it becomes exponentially unlikely for both verifiers to pass the verification protocol. By increasing the maximum noise tolerance of the protocol we increase the size of $\delta$, thereby allowing smaller sample sizes in the verification protocol, which increases the re-usability of coins. If we choose $n=4$, our scheme would be able to tolerate $16.6\%$ noise, and for $n=14$ it can tolerate up to $23\%$ noise. This concludes the proof of security against forging.

In the next section, we prove an upper bound on the error tolerance achievable for a general class of classical verification quantum money schemes, and show this bound limits to $25\%$ as the dimension of the underlying states is increased. This implies that our protocols are nearly optimal in terms of error tolerance. When proving this result, we assume only that the coin is a collection of quantum states each identified with a secret classical string, and that to verify the coin the holder must declare a number of single bit values which can be checked against the classical record.

%%%%%%%%%%%%%%%%%%%%%%%%%%%%%%%%%%%%%%%%%%%%%%%%%%%%%%%%%%%%%%%%%%%%%%%%%%%
%%%%%%%%%%%%%%%%%%%%%%%%%%%%%%%%%%%%%%%%%%%%%%%%%%%%%%%%%%%%%%%%%%%%%%%%%%%
%%%%%%%%%%%%%%%%%%%%%%%%%%%%%%%%%%%%%%%%%%%%%%%%%%%%%%%%%%%%%%%%%%%%%%%%%%%

\section{Maximum achievable noise tolerance} \label{sec:max}

Suppose we have a scheme in which the coin consists of many independently chosen $n$-dimensional pure quantum states, $\phi_x = |\phi_x\rangle\langle \phi_x |$, with $x\in X$ and where $x$ is a classical bit string chosen according to some probability distribution. To verify each state, the holder performs some POVM, $\mathcal{M}_x = \{M^{\text{cor}}_x, M^{\text{inc}}_x \}$, to ascertain one bit of information about each of the states used in the verification protocol. The bit values resulting from the measurement outcomes are checked against a classical record to verify whether the coin is genuine or not.
\begin{lemma}
For any quantum money scheme of the above type, the maximum tolerable noise, $e_{\text{max}}$, must be less than
\beq
e_{\text{max}} \leq \frac{1}{2} - \frac{1}{4}\frac{n+2}{n+1}.
\eeq
\end{lemma}
\textit{Proof.} We prove this by explicitly illustrating a strategy available to the adversary. The adversary holds the unknown state $\phi_x$, which lives in Hilbert space $\mathcal{H}$. She extends the state to $\phi_x \otimes \Phi$, where $\Phi = \frac{1}{n} \id_n$, and symmetrises the system. Specifically, she performs the mapping
\beq
\phi_x \otimes \Phi \rightarrow S_2 (\phi_x \otimes \Phi ) S_2,
\eeq
where $S_2$ is the projector onto $\mathcal{H}^2_+$, the symmetric subspace of $\mathcal{H}^{\otimes 2}$, and where the state on the right hand side is not normalised. The resulting normalised state of each clone is \cite{KW99}
\beq
\eta_x = v \phi_x + (1-v) \Phi,
\eeq
where $v := \frac{1}{2}\:\frac{n+2}{n+1}$. By the correctness requirement of quantum money schemes, an honest measurement on the correct state should always give a correct answer so that the coin is declared valid, i.e.
\beq \label{eq:correctness}
\trace(M^{\text{cor}}_x \phi_x) = 1.
\eeq
We further assume that, without access to the state $\phi_x$, the adversary has no information on $x$ and can do no better than to guess randomly. This means her probability of declaring a correct bit value is $1/2$, i.e.\footnote{Note that this assumption holds for all hidden matching quantum money schemes considered, and for any scheme in which the verification protocol involves declaring many single bit values which are later checked. Nevertheless, there may be protocols in which the verification protocol involves checking many $m$-bit outcomes, in which case the more reasonable assumption would be 
\begin{equation*}
\trace(M^{\text{cor}}_x \Phi) = 1/2^m.
\end{equation*}
To our knowledge such a scheme does not exist, but if higher error tolerance is desired our proof suggests looking into such schemes.}
\beq \label{eq:guess}
\trace(M^{\text{cor}}_x \Phi) = 1/2.
\eeq
Both honest verifiers hold the state $\eta_x$. Using Eqs. \eqref{eq:correctness} and \eqref{eq:guess}, the probability that an honest verifier gets a correct measurement outcome is
\beq
\trace(M^{\text{cor}}_x\eta_x) = v \trace(M^{\text{cor}}_x \phi_x) + (1-v) \trace(M^{\text{cor}}_x \Phi) = v + \frac{(1-v)}{2}.
\eeq
Expressing $v$ in terms of the dimension of the system shows that this strategy (which is always available to the adversary) leads to the honest verifiers finding an error rate of
\beq
e_{\text{max}} = \frac{1}{2} - \frac{1}{4}\frac{n+2}{n+1},
\eeq
and so for any such scheme to be secure an honest participant must expect an error rate less than $e_{\text{max}}$ in an honest run of the protocol.

Our analysis shows that for any scheme with $n = 4$ the tolerable noise is at most $20\%$, which complements our results in Section \ref{sec:security} where we described a protocol with $n=4$ which tolerated noise up to $16.6\%$. For $n=14$, the bound in this section shows that any such scheme has a noise tolerance of at most $23.3\%$. For $n=14$, our protocol can achieve an error tolerance of $23.03\%$, and so it is nearly optimal. As we increase the dimension of the quantum states used for the coins, the upper bound on the tolerable noise approaches $25\%$ which coincides with our conjecture for the tolerable noise in our protocols above. 

\begin{figure}[h]
\includegraphics[width=0.5\textwidth]{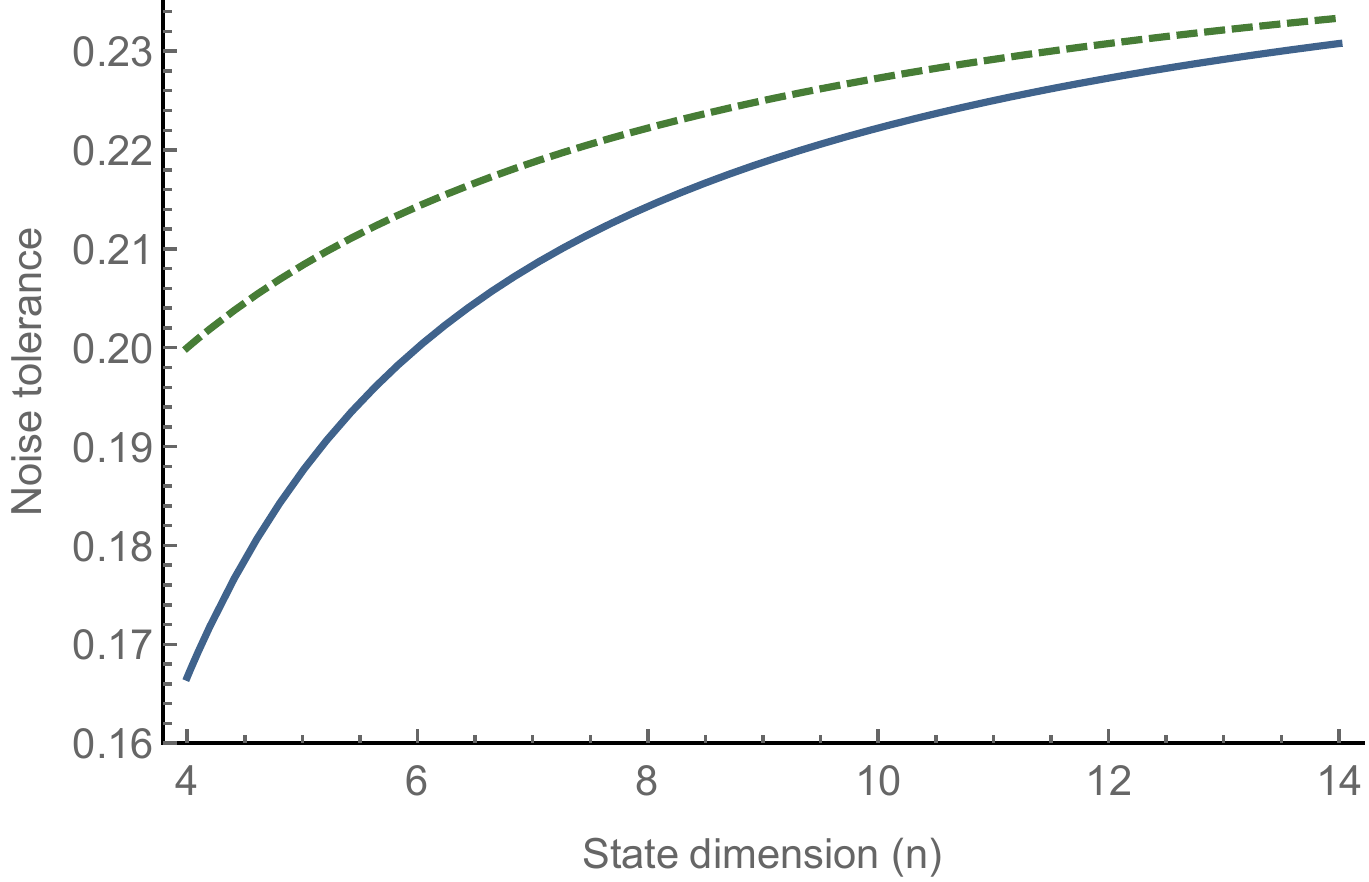}
\caption{Plot showing the theoretical bound on protocol noise tolerance (dotted line) and the noise tolerance achieved by the protocols in Section \ref{sec:scheme} (bold line) as the dimension of the underlying systems increase.}
\centering
\end{figure}

%%%%%%%%%%%%%%%%%%%%%%%%%%%%%%%%%%%%%%%%%%%%%%%%%%%%%%%%%%%%%%%%%%%%%%%%%%%
%%%%%%%%%%%%%%%%%%%%%%%%%%%%%%%%%%%%%%%%%%%%%%%%%%%%%%%%%%%%%%%%%%%%%%%%%%%
%%%%%%%%%%%%%%%%%%%%%%%%%%%%%%%%%%%%%%%%%%%%%%%%%%%%%%%%%%%%%%%%%%%%%%%%%%%

\section{Experimental Implementation}

The protocol presented in Section \ref{sec:scheme} gives rise to three main technical challenges when one considers experimental implementations, namely: the security analysis provided does not account for losses; the bank requires a source of complex, high-dimensional states; and the protocol requires that the coin holders have the ability to store states in quantum memory. In this section we address the first two issues so that a proof-of-principle implementation of the verification algorithm of the quantum money schemes could be performed with current technology.

\subsection{Detector Losses} \label{sec:detloss}

Here we tackle the first of the issues, and consider an implementation in which the verifiers use imperfect detectors with efficiency $\eta$. We assume that all detector losses are random and cannot be manipulated by the adversary. In this paper we do not consider channel loss, as we assume that coin transfers occur over short distances, meaning channel losses are less relevant. Nevertheless, many of the methods presented here would remain valid in the presence of small channel loss with only minor modifications necessary. To incorporate detector loss, it is necessary to modify the verification protocol, previously stated in Section \ref{sec:scheme}, so that it becomes:

\begin{algorithm}[H]
\floatname{algorithm}{Ver Algorithm}
\renewcommand{\thealgorithm}{}
\caption{}
\begin{algorithmic}[1]
\STATE The holder randomly chooses a subset of indices, $L\subset [q]$, with $l=|L|$, such that $r_i =0$ for each $i\in |$. The indices $i\in L$ specify the selection of games $G_i$ which will be used as tests for the verification procedure. For each $i\in L$, the holder then sets the corresponding bit of $r$ to be $1$ so that this game cannot be used in future verifications.
\STATE For each $i\in L$, the holder picks a relation $\sigma^\prime_i$ at random from $\mathcal{R}$ and applies the appropriate measurement to get answer $d_i$. If there is no measurement outcome we say the measurement was unsuccessful and set $d_i = \emptyset$. We define the number of successful measurement outcomes to be $l^\prime$. 
\STATE If $l^\prime < l_{min} := (\eta - \epsilon)l$, where $\epsilon > 0$ is a small security parameter, the verifier aborts the protocol.
\STATE The holder sends all triplets $(i, \sigma^\prime_i, d_i)$ to the bank.
\STATE The bank checks that $s<T$, where $T$ is the pre-defined maximum number of allowed verifications for the coin. If $s = T$, the bank declares the coin as invalid.
\STATE  For each $i$, the bank checks whether the answer is correct by comparing $(i, \sigma^\prime_i, d_i)$ to the secret $x^i$ values. The bank ignores those outcomes for which $d_i = \emptyset$, and accepts the coin as valid only if more than $l^\prime(c-\delta)$ of the answers are correct, where $c = 1-\beta$ is a measure of the channel correctness and $\delta$ is a small positive constant.
\STATE The bank updates $s$ to $s+1$.
\end{algorithmic}
\end{algorithm} 

\subsubsection{Correctness}
Correctness of the scheme follows from Hoeffding's inequality. When all participants are honest, it is exponentially unlikely for $l^\prime$ to be less than $l_{min}$, so the protocol will not abort, except with a negligible probability. If the protocol does not abort, the verifier has at least $l_{min}$ successful measurement outcomes, each with an independent probability $c$ of being correct. Overall, the probability of the verification failing is bounded by
\beq
\mathbb{P}(\text{Ver fails}) \leq \exp \left[-2l_{min}\delta^2\right] + \exp[-2l\epsilon^2],
\eeq
where now $\delta = (e^\prime_{\text{min}}-\beta)/2$, with $e^\prime_{\text{min}}$ derived in Eq. \eqref{eq:emin} below as the minimum average error rate achievable by the adversary.

\subsubsection{Unforgeability}

Since the protocol now includes detector losses, the adversary may not have to send states to each verifier for each game in the verification protocol, and she could attempt to hide losses arising from her strategy in the losses arising from detector inefficiency. As a consequence, the set of strategies available to the adversary is increased, and we must make sure our arguments in Section \ref{sec:security} still apply. 

Let $U_1$ and $U_2$ be $q$-bit strings representing whether or not the adversary sent a state to Ver$_1$ and Ver$_2$ respectively, for each of the $q$ games created by the bank. An entry of $1$ means the adversary sent a state to the verifier, while an entry of $0$ means the adversary did not send a state to the verifier. We want to show that, in order for the protocol not to abort, $W(U_i) \geq \gamma q$, where $\gamma := 1 - \frac{3\epsilon}{\eta}$ and $W$ is the Hamming weight. Suppose $W(U_i) = \gamma q$. Then, in Step 1 of the verification protocol, Ver$_i$ takes a sample, $V_i$, consisting of $l$ of the entries of $U_i$. Hoeffding's inequality gives
\beq
P\Big(W(V_i) \leq (\gamma+\frac{\epsilon}{\eta}) l\Big) \geq 1 - \exp [-2\frac{\epsilon^2}{\eta^2}l].
\eeq
If $W(V_i) \leq (\gamma+\frac{\epsilon}{\eta}) l$, then the probability of at least $l_{min}$ successful measurement outcomes is given by
\beq
P\Big(\text{At least $l_{min}$ successful measurement outcomes } | \:\:W(V_i) \leq (\gamma+\frac{\epsilon}{\eta}) l \Big) \leq \exp[-2l\epsilon^2] .
\eeq
The probability of the protocol proceeding past Step 3 of verification is therefore
\beq \label{eq:noabort}
P(\text{No Abort} | W(U_i) = \gamma q ) \leq \exp [-2\frac{\epsilon^2}{\eta^2}l]+ \exp[-2\epsilon^2l].
\eeq
In what follows we assume $W(U_i) \geq \gamma q$, since otherwise the above shows that the verifiers will abort with near certainty. This means the adversary is able to use any strategy that leads to channel losses of at most $\frac{3\epsilon}{\eta}$ for each verifier, as these can be hidden within the normal fluctuations of detector loss. Suppose there is a strategy which gives at least $(1-\frac{3\epsilon}{\eta}) q$ states to each verifier, and which leads to an average error probability (on only the states tested) of $e^\prime_{\text{min}}$ for at least one of the verifiers. Then, there is a strategy which gives $q$ states to each verifier, and leads to an average error probability for at least one of the verifiers of $(1-\frac{3\epsilon}{\eta})e^\prime_{\text{min}} + \frac{3\epsilon}{2\eta}$ (the adversary simply sends the maximally mixed state to each verifier in place of the $\frac{3\epsilon}{\eta}$ losses). Since this strategy falls under the scope of the analysis in Section \ref{sec:security}, we know that the resulting error rate must be at least $e_{\text{min}}$, which means
\beq \label{eq:emin}
e^\prime_{\text{min}} \geq \frac{e_{\text{min}} - \frac{3\epsilon}{2\eta}}{1-\frac{3\epsilon}{\eta}}.
\eeq
The parameter $\epsilon$ can be chosen to be arbitrarily small by increasing the sample size $l$. As such, the protocol is able to handle arbitrarily large detector losses, and leads to noise tolerance that can be kept arbitrarily close to the noise tolerance derived for the case of perfect detectors.

Each verifier tests at least $l_{min}$ states, and at least one verifier expects an error rate of $e^\prime_{\text{min}}$. The probability of this verifier passing the test is bounded as
\beq \label{eq:40}
P(\text{Observed error rate smaller than $e^\prime_{\text{min}} - \delta$}) \leq \exp [-2l_{min}\delta^2].
\eeq 
Combining Eqs. \eqref{eq:noabort} and \eqref{eq:40}, the probability that the adversary is able to forge a coin is given by
\beq
P(\text{Forgery}) \leq \exp [-2\frac{\epsilon^2}{\eta^2}l] + \exp [-2l\epsilon^2] + \exp [-2l_{min}\delta^2]
\eeq

%%%%%%%%%%%%%%%%%%%%%%%%%%%%%%%%%%%%%%%%%%%%%%%%%%%%%%%%%%%%%%%%%%%%%%%%%%%

\subsection{Coherent State Implementation}

In this section we tackle the second issue arising when considering experimental realisations of the scheme -- the bank must create hidden matching states of the form in Eq. \eqref{eq:HM}, which are high-dimensional states of high complexity. The implementation of hidden matching quantum retrieval games has been studied extensively in Ref. \cite{JM}, where the coherent state mapping defined in Ref. \cite{JM_Mapping} was used to approximate each hidden matching state by a sequence of $n$ coherent states of the form
\begin{align}
\ket{\alpha,x}&=e^{-\frac{|\alpha|^2}{2}}\sum_{k=0}^{\infty}\frac{\alpha^k}{k!}(a_x^{\dagger})^n\ket{0}\nonumber\\
&=\bigotimes_{i=1}^n\ket{(-1)^{x_i} \frac{\alpha}{\sqrt{n}}},
\end{align}
where
\beq
a_x^{\dagger}=\frac{1}{\sqrt{n}}\sum_{i=1}^{n}(-1)^{x_i}b_i^{\dagger}
\eeq
and $\{b_1^{\dagger},b_2^{\dagger},\ldots,b_n^{\dagger}\}$ are the creation operators of the $n$ modes. We call each sequence of coherent states a block, so that a single block is used to approximate a hidden matching state. As outlined in Ref. \cite{JM}, Bob's measurement can then be performed using linear optics circuits and single photon detectors.

In the absence of a phase reference, the phase of each block is randomised, which implies that each block is equivalent to a classical mixture of number states \cite{BLM2000}. More specifically, writing $\alpha=e^{i\theta}|\alpha|$, we have
\begin{align}
\int_{0}^{2\pi}\frac{d\theta}{2\pi}\ketbra{\alpha,x}{\alpha,x}=e^{-|\alpha|^2}\sum_{k=0}^{\infty}\frac{|\alpha|^{2k}}{k!}\ketbra{k}{k}_x,
\end{align}
where $\ketbra{k}{k}_x$ is a state of $k$ photons in the mode $a_x^{\dagger}$. Thus, the probability of obtaining a particular number of photons depends only on $\alpha$, which is a free parameter within the coherent state mapping. We consider the following three cases:

\subsubsection{Zero photons in the block}
In this case the state emitted is simply the vacuum state. If the adversary chooses to forward a state on to the verifiers, she can do no better than to induce a $50\%$ error rate, and it is simple to show that it is never beneficial for her to do so. This scenario can therefore be considered a ``source" loss, as opposed to a channel or detector loss. Crucially, since these losses are not controllable by the adversary, they can be treated in the same manner as detector losses in Section \ref{sec:detloss} simply by including the source loss into the detector loss parameter, $\eta$. The probability of zero photons being emitted is $p_0 = e^{-|\alpha|^2}$.

\subsubsection{One photon in the block}
In this case, the state emitted is equivalent to the ideal hidden matching state in Eq. \eqref{eq:HM} since
\begin{align}
\ket{1}_x&=a_x^{\dagger}\ket{0}\nonumber\\
&=\frac{1}{\sqrt{n}}\sum_{i=1}^{n}b_i^{\dagger}\ket{0}\nonumber\\
&=\frac{1}{\sqrt{n}}\sum_{i=1}^{n} (-1)^{x_i}\ket{i},
\end{align} 
where $\ket{i}$ is a single photon state in the mode $b_i$. Therefore, whenever the bank's source emits a single photon, the analysis in Section \ref{sec:security} applies. The probability of one photon being emitted is $p_1 = |\alpha|^2 e^{-|\alpha|^2}$.

\subsubsection{More than one photon in the block}
In this case we assume the worst case scenario: whenever the source emits more than one photon to represent a hidden matching state, the adversary can perfectly forge that state. The resulting error rate for the adversary is $e^\prime_{\text{min}}(\frac{p_1}{p_1+p_{2+}})$, where $p_{2+} = 1-p_0-p_1$. For small $|\alpha|$, $p_{2+}\approx \frac{|\alpha|^4}{2}$, while $p_1\approx |\alpha|^2$, so that $p_{2+}\ll p_1$ and the adversary's error probability is almost unchanged by using coherent states.

%%%%%%%%%%%%%%%%%%%%%%%%%%%%%%%%%%%%%%%%%%%%%%%%%%%%%%%%%%%%%%%%%%%%%%%%%%%
%%%%%%%%%%%%%%%%%%%%%%%%%%%%%%%%%%%%%%%%%%%%%%%%%%%%%%%%%%%%%%%%%%%%%%%%%%%
%%%%%%%%%%%%%%%%%%%%%%%%%%%%%%%%%%%%%%%%%%%%%%%%%%%%%%%%%%%%%%%%%%%%%%%%%%%

\section{Conclusion}

We presented a family of unconditionally secure classical verification quantum money schemes which are tolerant to noise up to $23\%$, and which we conjecture tolerate noise up to $25\%$. We further proved that $25\%$ is the maximum noise tolerance achievable for a wide class of quantum money schemes, including all classical verification secret-key schemes previously proposed. The security of our schemes depends on the difference between maximum tolerable noise and expected noise, meaning the increase in maximum tolerable noise increases the efficiency of our scheme, allowing for smaller, more re-usable coins. The techniques we use to prove security differ considerably to previous papers, and the re-usability of our coins is optimal \cite{Gav2012} in that it scales linearly with the number of qubits in the coin. This is a significant improvement when compared to Ref. \cite{GK2015}, in which the re-usability scales as $q^{1/3}$, and Ref. \cite{Gav2012}, in which re-usability scales as $q^{1/4}$, where $q$ is the total number of qubits in the coin. With realistic assumptions on experimental equipment, we expect that, using $n=8$, a coin containing $10^{9}$ qubits would use $l=18,000$ states for each verification, and would be re-usable $T = 100$ times for a security level of $10^{-6}$. Lastly, we suggested methods of adapting our techniques to facilitate experimental implementations of the scheme. We show that the schemes can be implemented using weak coherent states even in the presence of limited detector efficiency. 

%%%%%%%%%%%%%%%%%%%%%%%%%%%%%%%%%%%%%%%%%%%%%%%%%%%%%%%%%%%%%%%%%%%%%%%%%%%
%%%%%%%%%%%%%%%%%%%%%%%%%%%%%%%%%%%%%%%%%%%%%%%%%%%%%%%%%%%%%%%%%%%%%%%%%%%
%%%%%%%%%%%%%%%%%%%%%%%%%%%%%%%%%%%%%%%%%%%%%%%%%%%%%%%%%%%%%%%%%%%%%%%%%%%

\acknowledgements{The authors would like to thank I. Kerenidis, E. Andersson, and A. Ignjatovic for helpful discussions. R. A. gratefully acknowledges EPSRC studentship funding under grant number EP/I007002/1. J.M.A. recognizes funding from the Singapore Ministry of Education (partly through the Academic Research Fund Tier 3 MOE2012-T3-1-009) and the National Research Foundation of Singapore, Prime Minister’s Office, under the Research Centres of Excellence programme.}

%%%%%%%%%%%%%%%%%%%%%%%%%%%%%%%%%%%%%%%%%%%%%%%%%%%%%%%%%%%%%%%%%%%%%%%%%%%
%%%%%%%%%%%%%%%%%%%%%%%%%%%%%%%%%%%%%%%%%%%%%%%%%%%%%%%%%%%%%%%%%%%%%%%%%%%
%%%%%%%%%%%%%%%%%%%%%%%%%%%%%%%%%%%%%%%%%%%%%%%%%%%%%%%%%%%%%%%%%%%%%%%%%%%

\bibliography{QuantumMoneyBib}

\section{Appendix A}

\subsection{Overview of Argument}

In the main paper, we claim that the adversary cannot use coherent attacks on multiple states in order to beat the bound given in Eq. \eqref{eq:bound}, even when conditioned on the states chosen by the bank, and on the outcomes of previous measurement results found by the verifiers. In this section we formally prove our claim using a teleportation argument similar to the one introduced by Croke and Kent in Ref. \cite{CrokeKent}, so that each game can essentially be viewed as independent of all others.

In order to apply the teleportation argument, we must first introduce a modified individual setting, in which the adversary is allowed an additional ability. We show that this modification does not help the adversary to cheat. We then show that any coherent strategy can be transformed into a modified individual strategy. Therefore, any coherent strategy cannot beat the bounds proved for the unmodified individual case, as claimed.

\subsection{Modified Individual Attacks} \label{sec:indiv}

In the individual setting, the verifiers each receive a single hidden matching state and apply the verification protocol to test its authenticity. As specified by the protocol, the verifiers randomly choose to measure the state they receive using one of the matching measurements. We include this random choice of matching into the mathematical description of the measurement, and group the outcomes to be either ``correct" or ``incorrect". It can be shown that if the bank creates $\phi_x = \ket{\phi_x}\bra{\phi_x}$, the verifiers measurement is described by the POVM
\beq
\Gamma_x = \{ \Gamma^{cor, x}, \Gamma^{inc, x} \} = \frac{n}{2(n-1)} \left\{ \frac{n-2}{n} \mathbb{I} + \phi_x, \: \mathbb{I} - \phi_x  \right\}.
\eeq
Suppose now the adversary has the additional power of being able to force the verifiers to apply a correction unitary (which will be the teleportation corrections) to their measurement outcomes before they are sent to the bank. The adversary must specify the correction operation before sending the states to the verifiers, and, crucially, the correction operation is such that it is simply a permutation of the set of hidden matching states. For example, suppose the teleportation operation takes input $\ket{\phi_x}$ and outputs $\ket{\phi_{x^\prime}}$, with correction operator $C$. In this case, before sending the states, the adversary will tell the verifiers that they must apply correction $C$ to their measurement outcomes. In effect then, the verifiers will measure
\beq \label{eq:povm}
\Gamma_{x^\prime} = \{ \Gamma^{cor, x^\prime}, \Gamma^{inc, x^\prime} \} = \frac{n}{2(n-1)} \left\{ \frac{n-2}{n} \mathbb{I} + \phi_{x^\prime}, \: \mathbb{I} - \phi_{x^\prime}  \right\},
\eeq
since the correction applied to $\Gamma^{inc, x^\prime}$ is $\Gamma^{inc, x}$. On average, given $\phi_x$, it is not possible for the adversary to create two states, $\eta_x$ and $\tau_x$, such that $\trace[ \Gamma^{inc, x^\prime} (\eta_x + \tau_x) ] < p$. If it were possible, then it would imply that the adversary can clone $\phi_{x^\prime}$ better than what is allowed by quantum mechanics (and our arguments in the main paper). This is because if the adversary was given $\phi_{x^\prime}$ he could easily transform it to $\phi_x$ by applying $C$, and then perform the strategy to get two copies with a fidelity higher than the bound proved in the main paper. Therefore the additional power given to the adversary does not allow her to decrease the value of $p_{\text{Ver}_1} + p_{\text{Ver}_2}$.

\subsection{Coherent Strategy}

We now consider the case of $N$ games created by the bank. The bank creates
\beq
\frac{1}{2^{Nn}} \sum_{x_1,x_2} \ket{x_1}\bra{x_1}_{X_1} \otimes \ket{x_2}\bra{x_2}_{X_2}\otimes \ket{\phi_{x_1}}\bra{\phi_{x_1}}_A \otimes \ket{\phi_{x_2}}\bra{\phi_{x_2}}_B.
\eeq
The $X_1$ and $A$ registers contain the first $N-1$ secret strings selected by the bank and the corresponding hidden matching states, respectively. The $X_2$ and $B$ registers contain the $N$'th secret string selected by the bank and its corresponding hidden matching state. Only the $A$ and $B$ registers are accessible to the adversary. We assume for a contradiction that there exists a strategy available to the adversary such that, conditional on the value in the $X_1$ register, and conditional on the verifiers obtaining specific outcomes in previous measurements, the value of $p_{\text{Ver$_1$}} + p_{\text{Ver$_2$}}$ in the $N$'th game is decreased below the bound in Eq. \eqref{eq:bound}.

We describe this strategy as follows -- upon receiving the states from the bank, the adversary applies the unitary operation $S_{ABC}$ so that the state becomes
\beq
\begin{split}
& \frac{1}{2^{Nn}} \sum_{x_1,x_2} \ket{x_1}\bra{x_1}_{X_1} \otimes \ket{x_2}\bra{x_2}_{X_2}\otimes S_{ABC}\Big(\ket{\phi_{x_1}}\bra{\phi_{x_1}}_A \otimes \ket{\phi_{x_2}}\bra{\phi_{x_2}}_B \otimes \ket{0}\bra{0}_C \Big) S^\dagger_{ABC} \\
&= \frac{1}{2^{Nn}} \sum_{x_1,x_2} \ket{x_1}\bra{x_1}_{X_1} \otimes \ket{x_2}\bra{x_2}_{X_2}\otimes \ket{\Psi^{x_1x_2}}\bra{\Psi^{x_1x_2}}_{AA^\prime B B^\prime C^\prime}.
\end{split}
\eeq
The $A, A^\prime$ registers are the spaces that contain the states that will be sent to Ver$_1$ and Ver$_2$ (resp.) for the first $N-1$ games. The $B, B^\prime$ registers are the spaces that contain the states that will be sent to Ver$_1$ and Ver$_2$ (resp.) for the $N$'th game. The $C$ registers are auxiliary registers held by the adversary. We assume that the bank measures the $X_1$ register, and gets a state, $x_1$, which satisfies the conditions in the assumption. The state held by the adversary is then
\beq 
\frac{1}{2^{n}} \sum_{x_2} \ket{\Psi^{x_1x_2}} \bra{\Psi^{x_1x_2}}.
\eeq
The adversary gives the $A,A^\prime, B, B^\prime$ parts of the state to the verifiers. The honest verifiers will first make measurements on systems $A, A^\prime$ and a possible post measurement state is
\beq \label{eq:4}
\frac{1}{2^{n}} \sum_{x_2} a_{x_1x_2} \Pi_{AA^\prime}\ket{\Psi^{x_1x_2}} \bra{\Psi^{x_1x_2}} \Pi^\dagger_{AA^\prime}.
\eeq
We assume that $\Pi_{AA^\prime}$ is a measurement outcome satisfying the conditions of the assumption, so that the error probabilities on the $N$'th game are decreased. Here $a_{x_1x_2}$ is the normalisation term, $a_{x_1x_2} = 1/\trace \big[ \Pi_{AA^\prime}\ket{\Psi^{x_1x_2}}\bra{\Psi^{x_1x_2}}\Pi^\dagger_{AA^\prime}\big]$.

The verifiers now each measure $\Gamma_{x_2}$, as defined in Eq. \eqref{eq:povm}, on their $B$ system. The assumption tells us that 
\beq \label{eq:contr}
\frac{1}{2^{n}} \sum_{x_2} \Bigg[ a_{x_1x_2} \trace\Big[ \Gamma^{inc, x_2}_{B} \: \Pi_{AA^\prime}\ket{\Psi^{x_1x_2}} \bra{\Psi^{x_1x_2}} \Pi^\dagger_{AA^\prime} \Big] + a_{x_1x_2} \trace\Big[ \Gamma^{inc, x_2}_{B^\prime} \: \Pi_{AA^\prime}\ket{\Psi^{x_1x_2}} \bra{\Psi^{x_1x_2}} \Pi^\dagger_{AA^\prime} \Big] \Bigg] < p.
\eeq
We now aim to prove that this leads to a contradiction.

%%%%%%%%%%%%%%%%%%%%%%%%%%%%%%%%%%%%%%%%%%%%%%%%%%%%%%%%%%%%%%%%%%%%%%%%%%%%%%%%%%%%%%%%%

\subsection{Teleportation strategy}

Supposing the above strategy exists, we explore what this enables the adversary to do in the individual case in the hopes of finding a contradiction. We suppose the bank creates 
\beq
\frac{1}{2^{n}} \sum_{x_2} \ket{x_2}\bra{x_2}_{X_2} \otimes \ket{\phi_{x_2}}\bra{\phi_{x_2}}_B 
\eeq
and sends the $B$ part to the adversary. The adversary can simulate the above strategy locally, by creating $\ket{x_1}$, $\ket{\phi_{x_1}}$ and the maximally mixed state on $n$ dimensions $\ket{\Phi}$. After relabelling the registers, the adversary holds the state
\beq
\frac{1}{2^{n}} \sum_{x_2} \ket{x_1}\bra{x_1}_{X_1} \otimes \ket{x_2}\bra{x_2}_{X_2}\otimes \ket{\phi_{x_1}}\bra{\phi_{x_1}}_A \otimes \ket{\phi_{x_2}}\bra{\phi_{x_2}}_D  \otimes \ket{0}\bra{0}_C \otimes \ket{\Phi}\bra{\Phi}_{BE}.
\eeq
To simulate the strategy in the previous section, the adversary applies $S$ to the $A$, $B$ and $C$ registers, followed by a measurement on the resulting $A, A^\prime$ registers. Conditional on measurement outcome $\Pi_{AA^\prime}$, she then applies a generalised Bell measurement on the $D$ and $E$ registers in order to teleport the unknown state $\ket{\phi_{x_2}}$ into the $B$ register which was acted on by $S$ (modulo a teleportation correction). If the appropriate measurement outcome is not found, the adversary does not perform the Bell measurement and instead starts again. The resulting state is
\beq \label{eq:11}
\frac{1}{2^{n}} \sum_{x_2}  a_{x_1x^\prime_2} \Pi_{AA^\prime} \ket{\Psi^{x_1x^\prime_2}} \bra{\Psi^{x_1x^\prime_2}} \Pi^\dagger_{AA^\prime}.
\eeq
Notice the state contains $x^\prime_2$ since the Bell measurement does not faithfully teleport the state, and a correction is required which we have not performed. If the dimension of the hidden matching states is a power of two, the correction operators are simply tensor products of the Pauli operators \cite{Rig05}. Crucially, all corrections define a bijective mapping between $x^\prime_2$ and $x_2$, so that as $x_2$ cycles over all possible values so does $x^\prime_2$, and the probabilities are not affected (all corrections are equally likely, which must be the case so that information is not communicated faster than light). 

The state in Eq. \eqref{eq:11} is the same as the state in Eq. \eqref{eq:4}, but the measurements applied by the verifiers are correlated with the $X_2$ register held by the bank. Therefore, the verifiers failure probabilities are not the same when measuring the two states. Measurements on the state in Eq. \eqref{eq:4} leads to a failure probability of 
\beq \label{eq:12}
\frac{1}{2^{n}} \sum_{x_2} \Bigg[ a_{x_1x_2} \trace\Big[ \Gamma^{inc, x_2}_{B} \: \Pi_{AA^\prime}\ket{\Psi^{x_1x_2}} \bra{\Psi^{x_1x_2}} \Pi^\dagger_{AA^\prime} \Big] + a_{x_1x_2} \trace\Big[ \Gamma^{inc, x_2}_{B^\prime} \: \Pi_{AA^\prime}\ket{\Psi^{x_1x_2}} \bra{\Psi^{x_1x_2}} \Pi^\dagger_{AA^\prime} \Big] \Bigg],
\eeq
while measurements on the state in Eq. \eqref{eq:11} lead to a failure probability of
\beq \label{eq:13}
\frac{1}{2^{n}} \sum_{x_2}   \Bigg[ a_{x_1x^\prime_2} \trace\Big[ \Gamma^{inc, x_2}_{B} \: \Pi_{AA^\prime}\ket{\Psi^{x_1x^{\prime}_2}} \bra{\Psi^{x_1x^{\prime}_2}} \Pi^\dagger_{AA^\prime} \Big] + a_{x_1x^\prime_2} \trace\Big[ \Gamma^{inc, x_2}_{B^\prime} \: \Pi_{AA^\prime}\ket{\Psi^{x_1x^{\prime}_2}} \bra{\Psi^{x_1x^{\prime}_2}} \Pi^\dagger_{AA^\prime} \Big] \Bigg].
\eeq
The difference being the appearance of $x^\prime_2$ in the second expression. Nevertheless, the two can be made equal if the verifiers are forced to apply the teleportation correction unitary to their measurement outcomes. In effect, this correction relabels the measurement outcomes so that $\Gamma^{inc, x_2} \rightarrow \Gamma^{inc, x^\prime_2}$. Following this correction, the two expressions \eqref{eq:12} and \eqref{eq:13} are equal. This shows that the assumption in Eq. \eqref{eq:contr} leads to a contradiction, since it shows an individual attack in the modified scenario can achieve the same error probability as a coherent attack, and the error probabilities achievable in the modified individual scenario are the same as for the unmodified individual scenario.

\end{document}